\documentclass[aps,prl,reprint,showpacs,showkeys,unsortedaddress,superscriptaddress]{revtex4-1}
\usepackage{latexsym}
\usepackage{graphicx}

\begin{document}

\title{Colloidal aggregation in microgravity by critical Casimir forces}

\author{Sandra J. Veen}
\author{Oleg Antoniuk}
\author{Bart Weber}
\affiliation{Van der Waals Zeeman Institute, University of Amsterdam, Science Park 904, 1098 XH Amsterdam, The Netherlands}
\author{Marco Potenza}
\affiliation{Department of Physics, University of Milan, via Celoria 16,  I-20133 Milan, Italy}
\author{Stefano Mazzoni}
\affiliation{European Space Agency/ESTEC, Keplerlaan 1, 2200AG Noordwijk, The Netherlands}
\author{Peter Schall}
\author{Gerard H. Wegdam}
\affiliation{Van der Waals Zeeman Institute, University of Amsterdam, Science Park 904, 1098 XH Amsterdam, The Netherlands}

\date{\today}

\begin{abstract}
By using the critical Casimir force, we study the attractive strength dependent aggregation of colloids with and without gravity by means of Near Field scattering. Significant differences were seen between microgravity and ground experiments, both in the structure of the formed fractal aggregates as well as the kinetics of growth. Ground measurements are severely affected by sedimentation resulting in reaction limited behavior. In microgravity, a purely diffusive behavior is seen reflected both in the measured fractal dimensions for the aggregates as well as the power law behavior in the rate of growth. Formed aggregates become more open as the attractive strength increases.
 
\end{abstract}

\maketitle

The quantum and classical world may have different fundamentals, on occasion they do meet. In the case of the Casimir effect, the same principle can be used in both fields. In the quantum Casimir effect an attraction between two dielectric plates in vacuum arises due to the confinement of zero-point fluctuations of the electromagnetic field \cite{Milton2001}. Fisher and de Genne were the first to realize that this phenomenon could also occur in classical systems \cite{Fisher1978,*DeGennes1981}. They showed that a force arises between objects (e.g. colloids) in a dielectric medium due to the confinement of bulk (density or concentration) fluctuations. This effect becomes most significant near the solvent critical point, where the correlation length becomes of the order of the separation between the objects. Hence the term critical Casimir effect.

Since then several groups reported on the experimental detection and use of the critical Casimir effect \cite{Hertlein2008,Zvyagolskaya2011,Bonn2009,Guo2008,Soyka2008,Garcia1999,Ganshin2006}. In colloidal systems in particular it can be used as an experimentally convenient way to control interactions between colloidal particles, namely just by adjusting the temperature \cite{Zvyagolskaya2011,Bonn2009,Guo2008}. By making use of a simple model for charge stabilized colloids in a binary mixture the temperature dependent aggregation and aggregate breakup can even be explained quantitatively \cite{Bonn2009,*Guo2008}. The great advantage of this over more traditional means, as depletion interactions or salt induced flocculation, is that the interaction is continuously tunable, reversible and accurate. Moreover, the critical Casimir force adjusts itself by a change in temperature on a molecular time scale resulting in an effective potential. This renders the interaction comparable to atomic systems.

There are many theoretical models that describe aggregation depending on one of two limiting steps: the transport of particles towards the aggregate or the sticking of the particles at the aggregate surface. Diffusion limited aggregation (DLA) for instance, describes the formation of structures by which the building blocks move by diffusion alone and stick as soon as they touch \cite{Witten1983}. On the other hand, reaction limited aggregation (RLA) was simulated by incorporating a sticking probability in order to account for an activation barrier in the sticking of the particles \cite{Meakin1983a}. This was also studied experimentally by systematically changing the activation barrier in a system containing charged colloids \cite{Weitz1985}. In most experimental cases however, it is diffusion and reaction limited cluster aggregation that is observed \cite{Ball1987,*Lin1989,*Lin1990}. Pure DLA is not seen in colloidal systems because convection caused by gravity prevents it.

By performing experiments in microgravity, a purely diffusive behavior is ensured. Moreover, due to the nature of the critical Casimir force we are able to directly study the relation between the particle potential and the structures formed. A relation which up till now has not been investigated. We report here on the first systematic study of the attractive strength dependent aggregation of spherical colloids with and without gravity. A profound difference was seen in the rate of growth of aggregates and in their fractal dimension between microgravity and ground experiments. Furthermore, microgravity measurements revealed an attractive strength dependence of the aggregate fractal dimension. 

The microgravity experiment operated on board of the Columbus module of the International Space Station (ISS) in the Microgravity Science Glove box. The experiment could be controlled remotely from ESA's Spanish User Support and Operations Center in Madrid.
We used charge stabilized fluorinated latex particles with a radius of 200 nm, density of 1.6 g/mL and refractive index of 1.37, closely matched by that of the solvent mixture, 1.40. This "binary" mixture consists of 3-methyl pyridine (3MP) in water/heavy water with a 3MP weight fraction of \textsl{X$_{3MP}$}=0.39 and a D$_2$O/H$_2$O weight fraction of \textsl{X$_{hw}$}=0.63. This mixture has a lower critical solution temperature of 49$^\circ$C \cite{Prafulla1992}. Before addition of the colloids, the solvent mixture was purified by distillation under vacuum. Four different dispersions were prepared for which contact with oxygen was avoided. Each sample contained the same colloid volume fraction of  $\sim10^{-4}$ and four different salt concentrations of 0.31, 0.79, 1.5 and 2.7 mmol/L sodium chloride (NaCl). From these dispersions duplicate sets of samples were taken for microgravity and ground experiments.
The aggregation of the colloids was studied by means of Near Field Scattering (NFS)\cite{Ferri2004}. The samples were illuminated by a collimated laser beam 8 mm in diameter with a wavelength of 930 nm. The interference between the transmitted beam and the scattered light was recorded behind the sample, at a distance of 2.9 mm. A 20x microscope objective imaged this plane onto a CCD with a pixel size of 10 $\mu$m. Data analysis has been performed following the methods described in \cite{Ferri2004}. The data were corrected for a Talbot related effect according to reference \cite{Potenza2010}.
For each sample, the aggregation temperature, T$_{agg}$, was determined as the temperature at which aggregation was first observed within 20 minutes. We then investigated the aggregation at temperatures between T$_{agg}$ and T$_{agg}$ + 0.4$^\circ$, in steps of 0.1$^\circ$. According to previous calculations, an increase of 0.1$^\circ$C corresponds to an increase in the well depth and hence increase in attractive strength of roughly 3 $k_bT$ \cite{Bonn2009} ($k_b$ = Boltzmann constant, T = temperature). For each temperature, the aggregation process was followed for one hour by acquiring NFS images with a frame rate of 1 $s^{-1}$. The sample was then cooled to a temperature far below T$_{agg}$, followed by stirring for at least 3 hours before a new measurement was started.

\begin{figure}
	\includegraphics{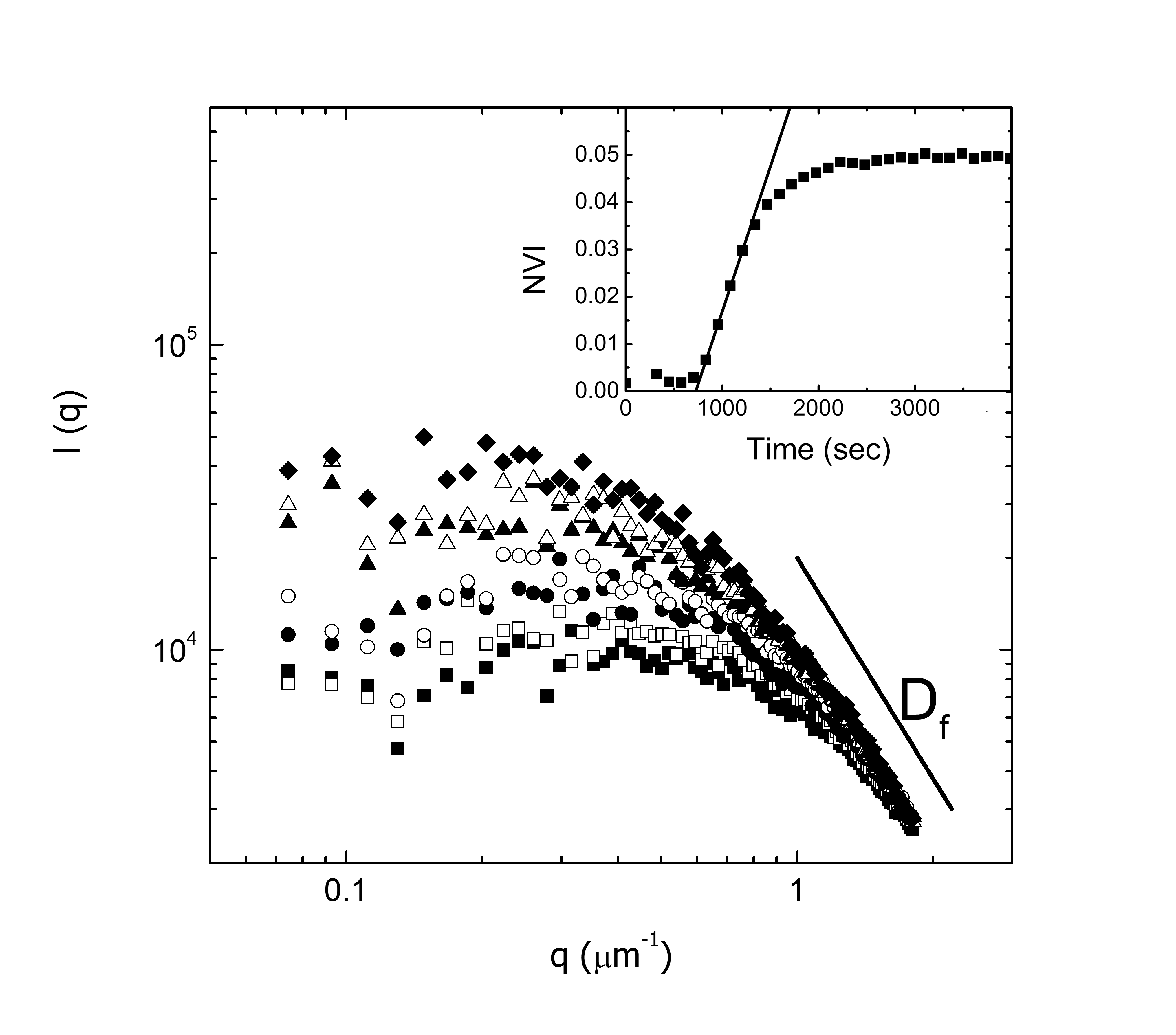}
	\caption{Typical light scattering result of the aggregation of colloids in microgravity. Different curves represent different points in time during the aggregation process with the overall scattering intensity increasing as the aggregates grow. Inset: Normalized variance of the scattering intensity (NVI) in time.(Sample: 0.31 mmol/L NaCl at T$_{agg}$)}
	\label{fig1}
\end{figure}

\begin{table*}
\caption{\label{table1}Fractal dimensions for different temperatures. For two samples (at the highest and lowest salt concentration used) the fractal dimension ($D_f$) for different temperatures is given as measured under microgravity. On ground, there was no dependence of $D_f$ on temperature, hence an average value is given.}
\begin{ruledtabular}
\begin{tabular}{c|cccccc}
Salt concentration & T$_{agg}$ & T$_{agg}$ + 0.1 & T$_{agg}$ + 0.2 & T$_{agg}$ + 0.3 & T$_{agg}$ + 0.4 & Ground \\
mmol/L & $D_f$ & $D_f$ & $D_f$ & $D_f$ & $D_f$ & Average $D_f$ \\
\hline
0.31 & 2.4 & 2.0 & 2.0 & 1.9 & 1.8 & 1.6 \\
2.7 & 2.4 & 2.1 & 2.1 & 1.8 & 2.0 & 1.8 \\
\end{tabular}
\end{ruledtabular}
\end{table*}

The growth of aggregates is reflected in the time evolution of the normalized variance of the scattered intensity (NVI) (inset Fig \ref{fig1}). This curve is the result of the combination of the changes in the number density of scatterers and their cross sections. The start of aggregation is marked by an increase in the NVI which is mainly due to the increase of the cross section. For small particles this depends strongly on the scatterer radius. In order to compare different measurements, the time at which aggregation starts had to be determined. A typical measured NVI curve shows a small bump, followed by a strong linear increase (inset Fig. \ref{fig1}). The start of aggregation, $t_0$, was determined by extrapolation of the linear regime to zero NVI. The initial bump was not taken into account because the temperature inside the cell was not yet constant. 

The NFS data allows us to elucidate the structure and average length scale of the aggregates as they grow. This is reflected in the angle dependent scattering intensity as shown in Figure \ref{fig1}. Here the scattering intensity is given as a function of the scattering wave vector $q$ ($q = (4\pi/\lambda) sin (\theta/2)$, in which $\lambda$ is the wavelength of the radiation used and $\theta$ is the scattering angle). The different symbols represent different times of growth with the overall scattering intensity increasing as the aggregates grow. For each sample a similar series of scattering curves was recorded at different temperatures at and above T$_{agg}$.

At high $q$ all distributions superimpose onto the same asymptotic curve which follows a power law behavior. This behavior is characteristic for scattering by fractal structures \cite{Hiemenz1997}. In Figure \ref{fig1} the intensity distributions are given on a log-log scale, converting the power law dependence into a linear form of which the slope equals the fractal dimension $D_f$. The fractal dimension is a measure for the internal structure of the aggregates and relates the radius of gyration ($R_g$) to the number of particles (\textsl{N}) by $R_g \propto N^{1/D_f}$. For a closed packed structure, $D_f$ would be close to 3.

We found that in microgravity the fractal dimension varied systematically with temperature controlling the attractive interactions (Table \ref{table1}). $D_f$ decreases as the temperature increases: from 2.4 at low temperatures, close to the theoretical DLA limit, to about 1.8 at high temperatures \cite{Meakin1983a}. The structures become more open as the attractive strength is increased. $D_f$ was constant throughout the measurement for the reported temperatures. Almost no dependence on the salt concentration and hence the repulsive part of the potential was seen. On ground there was no effect of the temperature on $D_f$ (Table \ref{table1}). Here the determined average $D_f$ was close to or lower then the highest temperature measurement in microgravity for the same sample.

\begin{figure}
	\includegraphics{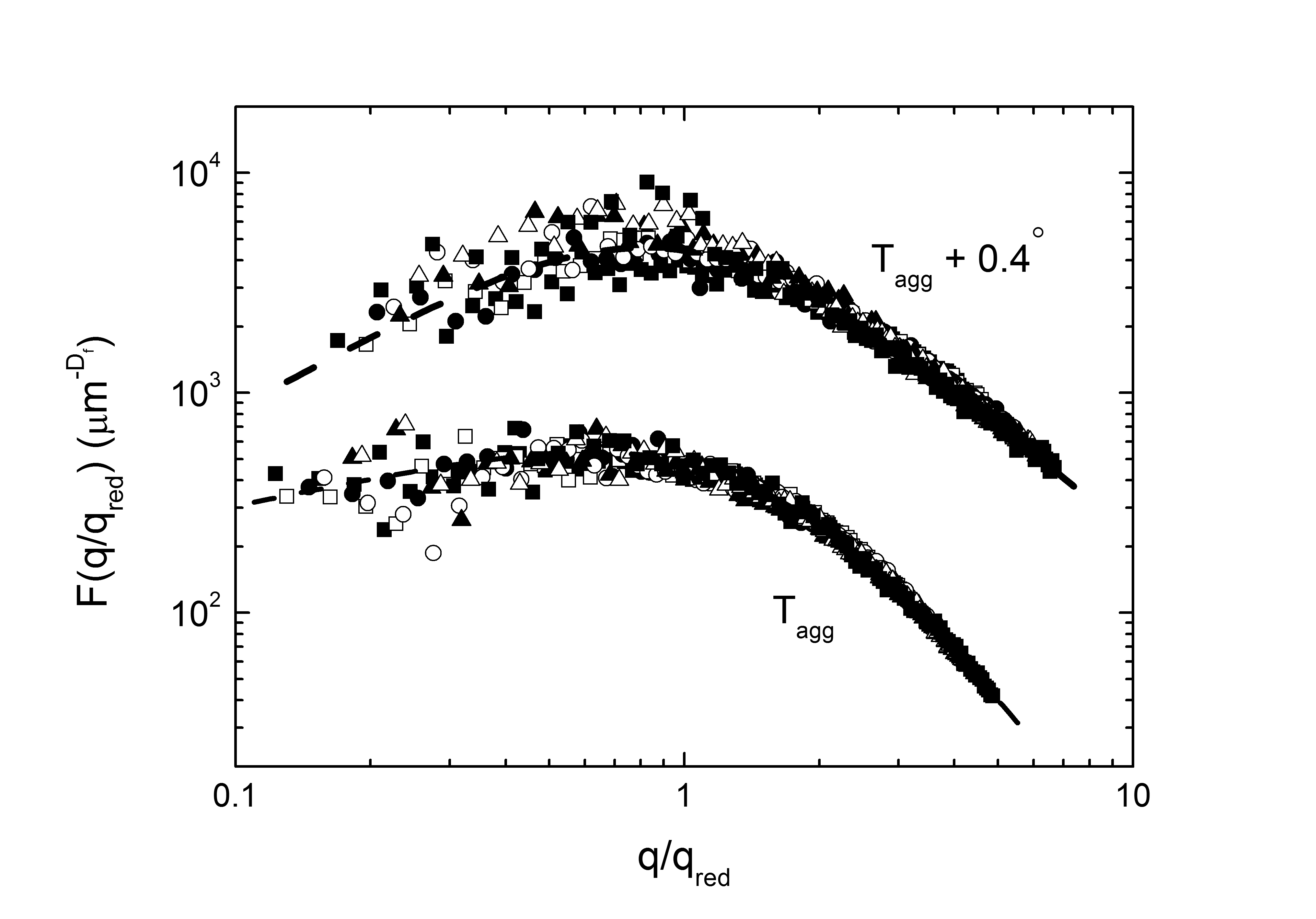}
	\caption{Collapse of scattering curves in microgravity. Rescaled scattering intensity as a function of rescaled wave vector \textsl{q} (Sample: 0.31 mmol/L NaCl). A perfect collapse is possible by taking into account the measured fractal dimension ($D_f$) and a reduced \textsl{q} value $q_{red}$. Lines are drawn as a guide to the eye in order to demonstrate the depression at low $q$ values.}
	\label{fig2}
\end{figure}

Despite the difference in attractive strength, the growth of aggregates at different temperatures does show similar behavior. As shown in Figure \ref{fig2} for two measurements in microgravity, the intensity distributions of the growing aggregates can be reduced to a single form using the measured $D_f$ and: 

\begin{equation}
S(q,q_{red}) = q_{red}^{-D_f} F(q/q_{red})
\label{eq1}
\end{equation}

This kind of collapse is characteristic for a type of spinodal decomposition process \cite{Gunton1972}. In regular spinodal decomposition the intensity distributions are scaled by using $q_{red}^{-D}$, with D being the system's dimensionality. The similar collapse of the measured intensity distributions suggests a possible common mechanism in the dynamics of this aggregation process \cite{Carpineti1992}. Experiments on the gelation of colloidal particles by means of depletion forces suggest that diffusion limited aggregation is in fact a deeply quenched limit of spinodal decomposition \cite{Lu2008}; a limit which is reached as soon as the interaction strength becomes much larger than the thermal energy. For our system this is indeed the case at the higher temperatures (theoretical strength ranges from ~3 to 15 k$_b$T \cite{Bonn2009}). Our results therefore substantiate this point of view.

By making use of the collapse of the intensity distributions, differences between the growth of aggregates in microgravity at different attractive strengths become more evident. For instance at T$_{agg}$ + 0.4$^\circ$, the highest measured attractive strength, a clear peak can be distinguished. In contrast, at the lowest attractive strength at T$_{agg}$, the maximum is not very pronounced (Figure \ref{fig2}). Scattering curves for spinodal decomposition contain a peak which moves towards lower $q$ in time indicating the growth of a characteristic length scale in the system. A minimum at low $q$ is characteristic for a region around the aggregate depleted of colloids \cite{Carpineti1992}. For the microgravity experiments, this peak and minimum are more pronounced at higher temperature. This indicates that when the interaction strength between the particles increases, so does the depletion region around the aggregates.

\begin{figure}
	\includegraphics{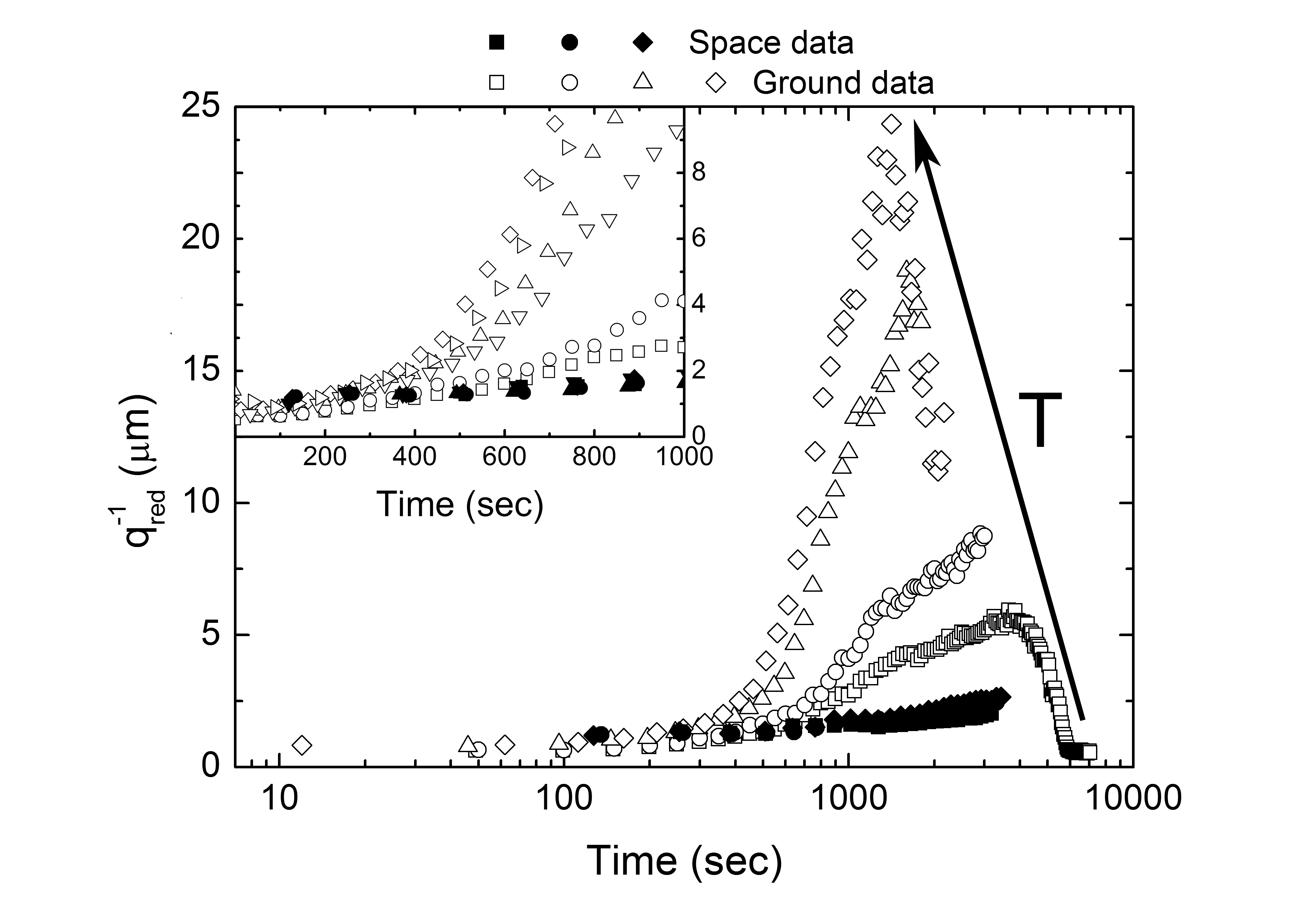}
	\caption{Growth of structures in time at different temperatures. Results from both microgravity (closed symbols) and ground experiments (open symbols) are shown (Sample: 1.5 mmol/L NaCl). The quantity $q_{red}^{-1}$ was obtained from the collapse of the intensity distributions and represents a typical length scale in the system. Inset: at the start of the aggregation process, both ground and microgravity measurements show the same growth speed.}
	\label{fig3}
\end{figure}

The kinetics or rate of growth of the aggregates offers insight into the underlying mechanism controlling the aggregate growth. A characteristic length scale of the system is given by $q_{red}^{-1} \propto R_g \propto N^{1/D_f}$ as determined by Eq. \ref{eq1}. The evolution of $q_{red}^{-1}$ in time for both the microgravity and ground experiments, corrected for the starting time $t_0$ as determined from the NVI, is given in Figure \ref{fig3}.

It is immediately evident that the speed of growth of the aggregates in microgravity is significantly lower than for the same experiments on ground. In microgravity, the aggregate size increases steadily in time.  This differs from what is seen for hard-sphere colloids where microgravity experiments revealed a two-step process in the rate of growth of their crystallites \cite{Zhu1997,*Cheng2002}. In our case, within the uncertainty of the determined $t_0$, the growth rate follows a power law with the exponent $1/D_f$ for all temperatures. This behavior is characteristic for diffusion limited models \cite{Meakin1984,Witten1983}. The fractal dimension at the lowest attractive strength is also close to the theoretical fractal dimension for DLA (2.5). These results show that microgravity indeed results in diffusion limited behavior. Surprisingly enough, DLA is already seen at the lowest interaction strength, pointing to a theoretical sticking probability of one. Increasing the attractive strength results in even more open structures and points to a change in the growth mechanism within the boundaries of the diffusion limited regime. This is something which has not been observed before experimentally.

On ground the behavior is clearly different. At early stages the growth rate of the aggregates follows the purely diffusive case as is seen in microgravity (see inset Figure \ref{fig3}). After this, there is a sharp and fast increase in the size of the aggregates in which the rate follows a more exponential form. In this regime, a strong dependence of the rate of growth on the temperature is observed. Aggregates grow faster as the attractive strength increases. Both the exponential form as well as the attractive strength dependence are consistent with a reaction limited model. The higher the attractive strength, the larger the probability a particle will attach to the growing structure (or lower the chance of detachment) and the faster the aggregates grow. Furthermore, a maximum is visible in the curve which shifts to larger sizes and shorter times for higher temperatures. This maximum is consistent with the start of sedimentation. 

In summary, we have shown the first results of the aggregation of charge stabilized spherical colloids due to critical Casimir forces in microgravity and on ground. Ground measurements are severely affected by sedimentation resulting in reaction limited behavior. The growth rate of aggregates strongly depends on the attractive strength. In microgravity, a purely diffusive behavior is seen reflected both in the measured fractal dimensions for the aggregates as well as the power law behavior in the rate of growth. Formed aggregates become more open as the attractive strength increases. The structure of the colloidal aggregates can thus be altered simply by changing the temperature, giving the opportunity to build specific structures in a controlled manner.

The authors kindly acknowledge M. Alaimo for his help on the ground measurements. S. Sacanna is thanked for the synthesis of the colloidal particles. This work was supported by the European Space Agency and the Dutch organization for scientific research NWO.

\bibliography{References}

\end{document}